\documentclass{elsarticle}
\usepackage{amssymb}
\usepackage{epsfig,graphicx}
\usepackage{graphicx}
\usepackage{amsmath}
\usepackage{color}
\usepackage{hyperref}
\hypersetup{
    colorlinks=true,
    linkcolor=blue,
    filecolor=magenta,      
    urlcolor=cyan,
    }
\newcommand{\be}{\begin{equation}}
\newcommand{\ee}{\end{equation}}

\newcommand{\bee}{\begin{eqnarray}}
\newcommand{\eee}{\end{eqnarray}}

\def\be{\begin{eqnarray} &&}

\def\ee{\end{eqnarray}}
\def\bew{\begin{widetext}}
\def\ew{\end{widetext}}

\usepackage{epsfig}

\def\bea{\begin{eqnarray}}
\def\eea{\end{eqnarray}}

\def\p {\pi}

\def\CP{CP }
\def\CPT{CPT }

\newcommand{\ppp}{\pi^{\pm} \pi^+ \pi^- }
\newcommand{\kkk}{K^{\pm} K^+ K^- }
\newcommand{\kkp}{\pi^{\pm} K^+ K^- }
\newcommand{\kpp}{K^{\pm} \pi^+ \pi^- }

\newcommand{\bppp}{$B^\pm \to \pi^\pm \pi^+\pi^- \,$}
\newcommand{\bkkk}{$B^\pm \to K^\pm K^+ K^- \,$}
\newcommand{\bkkp}{$B^\pm \to \pi^\pm K^+ K^- \,$}
\newcommand{\bkpp}{$B^\pm \to K^\pm \pi^+ \pi^- \,$}

\newcommand{\acp} {A_{cp}}

\begin{document}
% \makeatletter
\begin{frontmatter}
\title{Global CP asymmetries in charmless three-body $B$ decays  with final state interactions }
\author[CBPF]{I. Bediaga} 
\author[ITA]{T. Frederico}
%\author[ITA]{J. R. Lessa}
\author[ITA,UB]{P. C. Magalh\~aes}
\ead{p.magalhaes@bristol.ac.uk}
%\author[CBPF]{J. Miranda}
\author[CBPF]{D. Torres Machado}
\address[CBPF]{Centro Brasileiro de Pesquisas F\'isicas, 
%R. Dr. Xavier Sigaud, 
22.290-180 Rio de Janeiro, RJ, Brazil}
\address[ITA]{Instituto Tecnol\'ogico de
Aeron\'autica, 12.228-900 S\~ao Jos\'e dos Campos, SP,
Brazil.}
\address[UB]{H.H. Wills Physics Laboratory, University of Bristol,  Bristol, BS8 1TLB, United Kingdom.  }
\date{\today}

\begin{abstract}
 We propose a theoretical  framework to understand the observable  global charge-parity (CP) violation in  charmless three-body $B^\pm$ decays.
 The  decay amplitudes consider  the effects of the $\pi\pi \to K K $ rescattering treated within a CPT invariant framework together with the U-spin symmetry relation, $s\leftrightarrow d$, which results $\pi\leftrightarrow K$ in the  final state.
This approach applied to a two-channel model provides the magnitudes and signs of the ratios of the global CP asymmetries for  \bkpp, \bkkk, \bkkp, and \bppp decays,   qualitatively consistent with those obtained from the available experimental data. In addition, by considering the neutral channels, 
we predict the ratios for the global CP asymmetries for these decays. 
\end{abstract}
\end{frontmatter}

\section{Introduction}
There is a long-term discussion involving the source of the strong phase needed to generate direct violation of the charge-parity (CPV) symmetry in charmless B decays.
 Indeed, two interfering amplitudes with  weak and strong phases are necessary to produce the CPV. 
The weak phase comes from the CKM matrix, through the ``tree'' contribution at the quark level: $b \to u$ plus a $\bar u d $ (or $\bar u s$), producing the CKM phase $\gamma$. 
For the strong phase, however, there are two possible theoretical sources: one from the ``penguin'' contribution at the quark level and the other through a hadronic interactions between the final states. The polemic is about the relative importance of each one. 

When it comes to three-body charmless B decays, experimental results from LHCb collaboration \cite{LHCb-PAPER-2014-044} observed CPV for the four charged channels in $B\to hhh$, where $h$ are charged kaons or pions, with an intriguing distribution in the available phase-space.
 The QCD techniques are not enough to account for all observed CPV (see e.g. the recent review~\cite{Carla_ig}), which bring the FSI mechanism to the center of the debate.  Note that even QCD factorization approaches are including non-perturbative "long-distance" contributions to take into account hadronic strong phases in the CPV~problem~\cite{Beneke2016, Keri}.

An experimental technique to highlight the CP asymmetry directly from data, the Mirandizing approach~\cite{Mirandizing1,Mirandizing2} applied recently by the LHCb collaboration~\cite{LHCb-PAPER-2014-044,LHCb-PAPER-2013-027,LHCb-PAPER-2013-051}, showed a large variation of positive and negative  CP asymmetry in the Dalitz plot distribution. In particular in Ref.~\cite{LHCb-PAPER-2014-044}  this was shown to be up to 60$\%$ in specific regions in the Dalitz plane. Since the CKM weak phase must be independent of the position in the phase space, the change of sign must be directly related to the variation of the strong phase along the phase space.

Another aspect that should be considered in the  understanding of CP asymmetries  in $B$ decays is the so-called U-spin symmetry, which is a SU(2) subgroup of the SU(3) flavor, under which the $(d,s)$ pairs of quarks form a doublet, similar to $(u,d)$ isospin doublet~\cite{Lipkin96}.
The U-spin symmetry approach has been called to explain the observed CPV in charmless B decays~\cite{gronau95}, focused on the relation between decays channels with different strangeness quantum numbers. This approach succeed to reproduce the observed ratio between the CPV asymmetry in the $B^0_s \to K^-\pi^+$  and $B^0 \to K^+\pi^-$ partial widths~\cite{Carla_ig,gronau95}.

 Moving to three-body channels, Gronau and collaborators~\cite{gronau1,Bhattacharya:2013cvn} related the CPV asymmetries of the partial decay widths of the channels $B\to hhh$ based on the U-spin symmetry. 
 They pointed out  a relative minus sign between  \bkpp and \bkkp, as well as that between  \bkkk and \bppp.

To study the CP asymmetry in three-body B decays 
we use the difference between the partial decays widths of the charge conjugated states.
Such difference when summed up over all possible decay channels are constrained by CPT theorem to vanish \cite{Branco, bigi_book}. This difference 
is given by:
\begin{equation}\label{eq:DeltaGammaCP}
\Delta\Gamma_{\CP}(h_1^\pm h_2^+h_3^-)=\Gamma(B^- \to h_1^-h_2^+h_3^-) - \Gamma(B^+ \to h_1^+h_2^-h_3^+)\, .
\end{equation}
We can express $\Delta\Gamma_{\CP}$ from  the experimental integrated $A_{cp}$ results  through the equation:
\begin{equation}
\Delta\Gamma_{\CP}( h_1^\pm h_2^+ h_3^-) = A_{\CP}(B^\pm \to h_1^\pm h_2^+ h_3^-)\mathcal{B} (B^+\to h_1^+ h_2^+ h_3^-)/\tau(B^+).
\end{equation}
Where we used the experimental data quoted in~\cite{PDG} for the branching ratios $\mathcal{B}$, lifetime $\tau(B^+)$ and $A_{\CP}$. The resulting $\Delta\Gamma_{\CP}$ values are given in Table~\ref{W_cp}.

\begingroup
\begin{table}[htb] 
\caption{Width difference between the charge conjugate states $\Delta\Gamma_{\CP}$ for specific decays channels. Lifetime, branching ratios and $\acp$ are given as average PDG~\cite{PDG} values with the statistical and systematic uncertainties added in quadrature.}
\begin{center} 
\begin{tabular}{lc}
\hline 
 Decay channel & $\Delta\Gamma_{\CP} ( 10^{6} \, \textrm{s}^{-1})$ \\ %\midrule 
\hline 
\bkpp & $+0.84 \pm 0.25$ \\ 
\bkkk & $-0.68 \pm 0.17$ \\ 
\bppp & $+0.53 \pm 0.13$ \\ 
\bkkp & $-0.39 \pm 0.07$ \\ 
\hline 
\end{tabular} \end{center} 
\label{W_cp}
\end{table} 
\endgroup

The ratios between channels with different strangeness are:
\begin{equation}\label{eq:deltacpexp-a}
\frac{\Delta\Gamma_{\CP}(\pi^\pm K^+ K^-)}{\Delta\Gamma_{\CP}(K^\pm \p^+ \p^-)} = - 0.46 \pm 0.16\,\, \text{and}\,\, \frac{\Delta\Gamma_{\CP}(\pi^\pm \pi^+ \pi^-)}{\Delta\Gamma_{\CP}(K^\pm K^+ K^-)} = -0.77 \pm 0.27 \,,
\end{equation}
  which are compatible with -1  at 3$\sigma$ level for the first ratio and  1$\sigma$ for the second one.
These values are qualitatively consistent with U-spin symmetry as predicted by \cite{Bhattacharya:2013cvn}.
 
Exploring the possible final state interactions between the four charged channels and imposing the CPT constraint, we showed in \cite{BediagaPRD2014,Nogueira2015}  that $\pi\pi \to KK$ rescattering amplitude can explain the flip in the sign for $A_{cp}$ between channels coupled by the strong interaction, i.e. have the same strangeness. 
 Furthermore, from Table~\ref{W_cp}, we can get the complementary ratios: 
\begin{equation}\label{eq:deltacpexp-b}
\frac{\Delta\Gamma_{\CP}(\kpp)}{\Delta\Gamma_{\CP}(\ppp)} = 1.59 \pm 0.62 \,\text{ and }\, \frac{\Delta\Gamma_{\CP}(\kkk)}{\Delta\Gamma_{\CP}(\kkp)} = 1.77 \pm 0.55,
\end{equation} 
 which will be understood when we include the FSI contribution from $ \pi\pi \to KK$  together with the  U-spin symmetry of  the final state. 
  Before, these ratios were considered a puzzle and associated  to the U-spin symmetry breaking~\cite{chineses}.
  
Returning to the observable $\Delta\Gamma_{CP}$ in Table~\ref{W_cp}, 
from one side we have a qualitative agreement of U-spin prediction for the two ratios between channels with different strangeness.
And, on the other side, from the hadronic FSI approach, one can understand the relative sign between the two pairs of channels coupled through $\pi\pi\leftrightarrow KK$ interaction. 

A complete understanding of the observables in Table \ref{W_cp} is not trivial. We are dealing with three-body final states and one has to consider the complexity of their dynamics, with each channel being produced through several different intermediate states with different interference between them.  The global CP asymmetries result from all these dynamical contributions after integration of the differential decay rates over the phase space. With this perspective, our first task is to understand the signs and the modulus (around unity) of all the ratios in Table \ref{W_cp}. We remind that, to make the situation even more challenging, the channels have a different branching fraction, e.g. the \bkpp is one order of magnitude larger than the \bkkp.

Our work unifies two general frameworks to study the total CP violation related to charmless three-body B decays: the CP asymmetry associated with the U-spin approach  and the central role of hadronic final state interactions in these decays within the constraint of CPT invariance.
We go beyond previous works that used U-spin symmetry
  by considering final state interactions.

\section{Hints of FSI on data}
\label{hfsi}
 The rescattering process can be the source of strong phase and absorptive contributions in multi-body decays through the strong interaction   including also loops. 
This idea was proposed by Wolfenstein~\cite{wolfenstein} many years ago, and further investigated in several studies~\cite{Donoghue1996, Buras1997,AtwoodSoni1997,Suzuki2007,Soni2005,Smith2003,pat-mane2020}, including the CPV on B decays~\cite{BediagaPRD2014,Nogueira2015,ITP_2020, ITP_B3K,London1, London_gronau, Blok} (for more references see the review~\cite{Carla_ig}). In particular, in Refs.~\cite{BediagaPRD2014,Nogueira2015,ITP_2020, ITP_B3K} we discussed the relevance of CPT and rescattering as a mechanism of CPV in B three-body decays in light of the LHCb experimental results.

We should remind that, in the QCD-only approach, based on the BSS model~\cite{BSS}, the imaginary part from the strong interaction appears in the ``penguin'' diagram $b \to s$ (or d), plus $u \bar u$ or $d \bar d$ produced by the presence of an intermediary gluon. However, this occurs when it has transferred momentum twice the charm quark mass present in the ``penguin'' loop. While in models including hadronic rescattering, the strong phase can also be originated from process characterized by long distance physics~\cite{Nogueira2015,ITP_2020}.

Two-body scattering data  was measured  both  for $\p\p \to \p\p$ and  for $\p\p \to KK$ processes~\cite{CERN-Munich,Cohen1980}. The data suggest a strong coupling between these two channels in the S-wave.  They can also be coupled in P and D-waves, but data show that these couplings are  very small~\cite{PDG}. On the theory side, there are several parametrizations and  theoretical models that describe  the S-wave data well up to a certain energy (1.9 GeV)~\cite{Pelaez-Rodas, Pelaez06,  pat-mane2020} (and references included). 
In particular, the one from \cite{Pelaez06} was used to introduce the rescattering $\pi\pi \to KK$ S-wave amplitude in \bppp and \bkkp analysis at LHCb~\cite{LHCbPRL2020,LHCbPRD2020,LHCb-PAPER-2018-051}. 

To stress the relevance of FSI to the CPV observed in data, we  show in  Table~\ref{ACP_rescattering} the CP asymmetry from the rescattering $\pi\pi\to KK$ region  of the Dalitz plane ($A^{par}_{\CP}$) - from $1$ to $1.5 $~GeV -  along with the total $A_{CP}$ for the charmless charged three-body B decays: \bkpp, \bkkk, \bkkp, and \bppp.

\begingroup
\begin{table}[htb] 
\caption{Total charge asymmetries $A^{all}_{\CP}$ and partial ones $A^{Par}_{\CP}$ in the rescattering region $\pi\pi \to KK$ from 1.0 up to 1.5~GeV/c$^2$. Uncertainties are only statistical \cite{LHCb-PAPER-2014-044}.}
\begin{center} 
\begin{tabular}{lcc}
\hline 
  Decay & $A^{all}_{\CP}$ & $A^{par}_{\CP}$\\ %\midrule 
\hline 
\bkpp & $+0.025 \pm 0.004$ & $+0.123 \pm 0.012$ \\ 
\bkkk & $-0.036 \pm 0.004$ & $-0.209 \pm 0.011$ \\ 
\bppp & $+0.058 \pm 0.008$ & $+0.173 \pm 0.021$ \\ 
\bkkp & $-0.123 \pm 0.017$ & $-0.326 \pm 0.028$ \\ 
\hline 
\end{tabular} \end{center} 
\label{ACP_rescattering}
\end{table} 
\endgroup

The $\pi\pi\to KK$ rescattering as a source of CPV were investigated in a recent amplitude analysis performed by the LHCb collaboration~\cite{LHCbPRL2020,LHCbPRD2020,LHCb-PAPER-2018-051}, as we mentioned above.
The experimental result on the \bkkp decay~\cite{LHCb-PAPER-2018-051} shows a strong CP asymmetry associated with hadronic rescattering amplitude $\pi\pi \to KK$. The observed $A_{cp}= -66.4 \pm 3.8 \pm 1.9 \%$ represents the most significant CPV observed in a single amplitude. It has a fit fraction of $(16.4 \pm 0.8 \pm 1.0)\%$ which results in a $(-10.9 \pm 0.8 \pm 0.7)\%$ contribution to the integrated CP asymmetry. It corresponds to almost the total integrated asymmetry ($\acp$(\bkkp)$= -0.123 \pm 0.017)$.
We can do the same exercise for the \bppp decay with the recent amplitude analysis published by LHCb~\cite{LHCbPRD2020}, where the contributions from $\sigma$ and $f_2(1270)$ represent roughly  all integrated asymmetry observed in the \bppp channel.

 To complement the above discussion, we mention that, in the amplitude  analysis LHCb performed~\cite{LHCbPRL2020,LHCbPRD2020,LHCb-PAPER-2018-051} it was shown  that some local contributions to the CPV appearing in the Dalitz plot disappeared after  integrating over the phase-space. This was the case for the interference between S and P waves around the $\rho$ resonance.  However, the amplitude analysis showed that this is not the case in the kinematic region where the  rescattering $\pi\pi\to KK$ is relevant. The contribution to CPV in this region not only survives the integration but it gives the dominant contribution to the $B^\pm\to \pi^\pm K^+K^-$ decay, as we  have pointed out quantitatively and in Table~\ref{ACP_rescattering}.

\section{U-spin approach for  $B\to hhh$ decays }\label{sec:uspin}

\begin{figure}[!htb]
  \includegraphics[width=10cm]{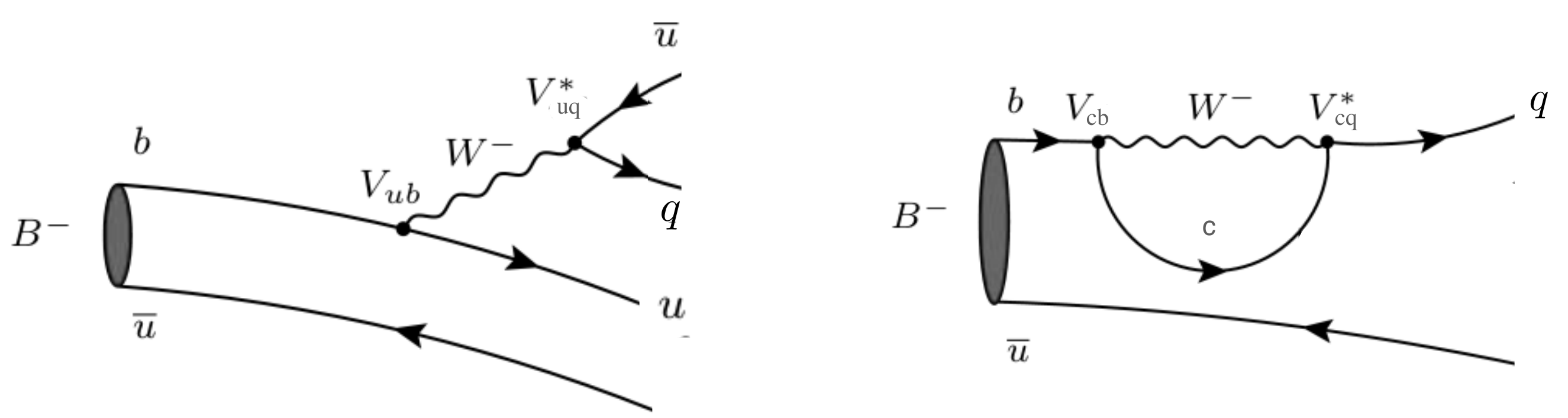}
  \caption{ All the four $B\to hhh$ channels  can have contribution from the ``tree'' (left) and ``penguin'' (right) diagrams. Note that we omit the gluon lines from the ``penguin''. The flavour of quark q can be  d or s, and the others quarks needed to the final hadronic state are produced from the vacuum.}
  \label{fig:heff}
\end{figure}

 The $B\to hhh$, for $h=\p, K$, amplitude can be generically represented by the Feynman diagrams in Figure~\ref{fig:heff}, where we omit the gluon lines and the other quarks produced from the sea to complete the final state.  
 Implementing U-spin approach inspired in \cite{gronau1} and considering the two main topologies with different quark flavor transitions (Figure~ \ref{fig:heff}), the amplitude of $B\to f $ decays, for $f=hhh$  ($f$ implicitly denotes the  momentum dependence associated with a point in the Dalitz plot), are given by: 
 \begin{eqnarray}\label{cpt3eff0}
 A(B^u \to f^q)=\langle f^q_{out}
|\mathcal{H}_{\text{w}}| B^u  \rangle = V_{ub}V_{uq}^*\langle   f^q_{out}|U^{ q} |B^u\rangle + V_{cb}V^ *_{cq}\langle  f^q_{out}|C^{ q}| B^u\rangle \, ,
\end{eqnarray}  
and for the decay of the charge conjugate state:
\begin{eqnarray}\label{eq:barBbarf0}
A(\bar{B^u}\to \bar f^q)=\langle\bar f^ q_{out}
|\mathcal{H}_{\text{w}}|\bar{B^u}  \rangle=V^*_{ub}V_{uq}\langle \bar f^q_{out}| \bar U^q | \bar{B^u}\rangle 
 + V^*_{cb}V_{cq}\langle \bar  f^q_{out}| \bar C^q | \bar{B^u}\rangle \, ,
 \end{eqnarray}
where $q=s$ or $d$, namely channels with $\Delta S=1$ or 0, respectively.
 The effective Hamiltonian for the decay is written as $\mathcal{H}_{\text{w}}$, and the decay amplitude is separated with the matrix elements of operators $U^q$ and $C^q$, associated respectively with the ``tree'' (left panel) and ``penguin'' (right panel) diagrams of Figure~\ref{fig:heff}, and within our assumption do not contain the strong phase.
The strong phase in the decay amplitudes, Eq.~\eqref{cpt3eff0} and \eqref{eq:barBbarf0}, comes from $|f^q_{out}\rangle$ and its charge conjugate state, which are the scattering eigenstates of  the strong Hamiltonian.
 To complement, in our notation, the states $ |f^q\rangle$ are hadronic-free states, while $ |f^q_{out(in)}\rangle$ includes the distortion  due to the hadronic FSI.
 %, 
In principle, such separation is possible in general scattering theory, and it will be necessary when analyzing the Charge-Parity-Time reversal (CPT) symmetry constraint.

The $B$ decay amplitudes for channels with $\Delta S =0$, \bppp and \bkkp, correspond to  $q=d$ in Eqs.~\eqref{cpt3eff0} and \eqref{eq:barBbarf0}. In the case of $\Delta S =1$, the decay amplitudes for \bkpp and \bkkk are associated to $q=s$.

 The U-spin symmetry corresponds to the invariance of decay amplitudes upon  the exchange of the  light flavored quarks, $d\leftrightarrow s$ in all hadrons at the decay channel,  which in our notation is written as:
\begin{equation}\label{h-hbaruspin}
\langle  f^s_{out}| U^s |B^u\rangle=\langle  f^d_{out}| U^d | B^u\rangle 
\quad\text{and}\quad \langle  f^s_{out}|C^s |B^u\rangle=\langle  f^d_{out}|C^d | B^u\rangle  \, .
\end{equation}
To further simplify the notation we define:
\begin{eqnarray}\label{auc}
\mathcal{U}_{f^q}=
\langle  f^q_{out}| U^q |B^u \rangle \quad \text{and} \quad
\mathcal{C}_{f^q} = \langle   f^q_{out}| C^q | B^u\rangle \, .
\end{eqnarray} 
 Note that  $|f^d_{out}\rangle$ and  $|f^s_{out}\rangle$ are  related by  the exchange of $\pi\leftrightarrow K$ in the final state used to compute the matrix elements appearing in the decay amplitude.
 
Considering the two main quark flavor topologies as in Figure~\ref{fig:heff}, the amplitudes corresponding to the charmless $B\to hhh $ decays are written as:
\begin{eqnarray}
\label{eq:amplitude_i}	
A(B^+ \to K^+ \pi^+\pi^-) = V_{ub}^* V_{us}\,\mathcal{U}_{s_1} + V_{cb}^* V_{cs}\,\mathcal{C}_{s_1},\\
\label{eq:amplitude_ii}
A(B^+ \to \pi^+ K^+ K^-) = V_{ub}^* V_{ud}\,\mathcal{U}_{d_2} + V_{cb}^* V_{cd}\, \mathcal{C}_{d_2},  \\
\label{eq:amplitude_iii}	
A(B^+ \to \pi^+ \pi^+\pi^-) = V_{ub}^* V_{ud}\,\mathcal{U}_{d_3} + V_{cb}^* V_{cd}\, \mathcal{C}_{d_3},\\
\label{eq:amplitude_iv}
A(B^+ \to K^+ K^+ K^-) = V_{ub}^* V_{us}\,\mathcal{U}_{s_4} + V_{cb}^* V_{cs} \,\mathcal{C}_{s_4}\, ,
\end{eqnarray}
where we have the channels $f^s=(s_1,\, s_4)$ and $f^d=(d_2,\, d_3)$.
The CP asymmetries in the partial widths, $\Delta\Gamma_{CP}(f)$  given  by Eq.~\eqref{eq:DeltaGammaCP}, 
comes from the interference terms in  $B\to f$ decays with different weak and strong phases, and by
considering the decay amplitudes \eqref{eq:amplitude_i}-\eqref{eq:amplitude_iv}, one arrives at:
\begin{eqnarray}
\label{eq:deltaAkpipi}	
 \Delta\Gamma_{CP}(K^\pm \pi^+\pi^-) &=& 2\,\text{Im}[V_{ub}^* V_{us}V_{cb}V_{cs}^* ]\, \text{Im}[\mathcal{U}_{s_1} \mathcal{C}_{s_1}^*
 +\bar{\mathcal U}_{s_1} 
 \bar {\mathcal C}^*_{s_1}],
 \\
\label{eq:deltaApikk}
\Delta\Gamma_{CP}( \pi^\pm K^+ K^-) &=& 2\,\text{Im}[V_{ub}^* V_{ud}V_{cb} V_{cd}^*]\, \text{Im}[\mathcal{U}_{d_2}\mathcal{C}^*_{d_2}+\bar{\mathcal U}_{d_2}\bar{\mathcal C}^*_{d_2}], \\
\label{eq:deltaApipipi}	
\Delta\Gamma_{CP}(\pi^\pm \pi^+ \pi^-) &=& 2\,\text{Im}[ V_{ub}^* V_{ud}V_{cb} V^*_{cd}]\,\text{Im}[\mathcal{U}_{d_3} \mathcal{C}_{d_3}^*
 +\bar{\mathcal U}_{d_3} 
 \bar {\mathcal C}^*_{d_3}],\\
\label{eq:deltaAkkk}
\Delta\Gamma_{CP}( K^\pm K^+ K^-) &=& 2\,\text{Im}[ V_{ub}^* V_{us}V_{cb} V^*_{cs}]\,\text{Im}[\mathcal{U}_{s_4}\mathcal{C}^*_{s_4}+\bar{\mathcal U}_{s_4}\bar{\mathcal C}^*_{s_4}]\, . 
\end{eqnarray} Imposing U-spin symmetry, expressed by Eq.~\eqref{h-hbaruspin}, one needs to make $d\leftrightarrow s$ in all mesons in the decay channel, namely:
\begin{equation}
  \mathcal{U}_{s_1}=\mathcal{U}_{d_2}\,, \quad \mathcal{C}_{s_1} =\mathcal{C}_{d_2}\, , \quad \mathcal{U}_{d_3}=\mathcal{U}_{s_4}\, , \quad \mathcal{C}_{d_3} =\mathcal{C}_{s_4}\, ,  
\end{equation} and considering that the unitarity of the CKM matrix leads to~\cite{gronau1,Bhattacharya:2013cvn}:
\begin{equation} \label{eq:uspin0-1}
\operatorname{Im}(V^*_{ub}V_{us}V_{cb}V^*_{cs}) = -\operatorname{Im}(V^*_{ub}V_{ud}V_{cb}V^*_{cd})\, ,
\end{equation} 
it can be shown that~\cite{gronau1,Bhattacharya:2013cvn}:
\bea
\Delta\Gamma_{CP}(K^\pm \pi^+\pi^-)=- \Delta\Gamma_{CP}( \pi^\pm K^+ K^-)\, ,\,\, 
\nonumber \\
\Delta\Gamma_{CP}(\pi^\pm \pi^+ \pi^-)=-
 \Delta\Gamma_{CP}( K^\pm K^+ K^-) \,.
\eea
These relations are  qualitatively consistent with the experimental results   within error given in Eq.~\eqref{eq:deltacpexp-a}. Still, it is  remaining the relation between the other observed width asymmetries given in Eq.~\eqref{eq:deltacpexp-b} and not only those related to the U-spin symmetry. For that purpose the CPT constraint in channels coupled by the strong interaction is necessary.

\section{FSI, U-spin symmetry and CPT}\label{sec:CPT}

As we discussed before, rescattering $\pi\pi\leftrightarrow KK$ can be a CPV mechanism in $B\to hhh$~\cite{BediagaPRD2014, Nogueira2015}. 
However, the question is how to connect the FSI between channels with the same quantum numbers with U-spin symmetry, that can only relate channels that have different $\Delta S$. Both are constrained by CPT theorem and all together should give a consistent description that allows us to understand the observable asymmetries in Table~\ref{W_cp}. 

First, to connect FSI with the observed CPV in B decays, we have to show that the relations given by Eqs.~\eqref{eq:deltaAkpipi} --  \eqref{eq:deltaAkkk} are consistent with the FSI formalism previously developed.
To define our notation and the FSI framework we follow the discussion presented in \cite{BediagaPRD2014} for implementing the \CPT constraint in $B$ meson decays, as developed in Refs.~\cite{Branco,Marshak}.

A hadron state $|h\rangle$ transforms under \CPT as ${\mathcal {CPT} }\,|h\rangle=\chi\langle \bar h|$, where $\bar h$ is the charge conjugate state, and $\chi$ is a phase. The weak and strong Hamiltonians are invariant under  CPT, and therefore it is valid that
$$({\mathcal {CPT}})^{-1}\,\mathcal{H}_{\text{w}} \,{\mathcal {CPT} }= \mathcal{H}_{\text{w}}\quad  \text{and} \quad 
({\mathcal {CPT} })^{-1}\,\mathcal{H}_s\,{\mathcal {CPT} }= \mathcal{H}_s\, . $$
%respectively.
%
The requirement of \CPT invariance for the weak and strong Hamiltonians imply that the sum of the partial decay widths of the hadron decays and the correspondent sum for the charge conjugate ones should be identical: 
\begin{eqnarray}
\sum_{f^q, \, q=d,s}|\langle f^q_{out}|\mathcal{H}_{\text{w}}|h\rangle|^2&=&
\sum_{f^q, \, q=d,s}
|\langle \bar f^q_{out} |\mathcal{H}_{\text{w}}| \bar h \rangle|^2\, .
\label{cpt5a}
\end{eqnarray} 
 We recall that in order to obtain the \CP asymmetry one has to take into account the change due to \CP transformation coming from the sign difference multiplying the weak phase.   The CP asymmetry formula that summarizes  Eqs.~\eqref{eq:deltaAkpipi}-\eqref{eq:deltaAkkk} is given by: 
\begin{eqnarray}\label{cpuspin}
\Delta\Gamma_{CP} (f^q) &=&|A(B^u\to f^q)|^2-|A(\bar {B^u}\to\bar f^ q)|^2
\nonumber \\ &=&
  2\,\text{Im}[V_{ub}^*V_{uq}V_{cb}V^*_{cq}]\,\text{Im}
 \left[\mathcal{U}_{f^q}
\mathcal{C}_{f^q}^*+\bar{\mathcal{U}}_{f^q}
\bar{\mathcal{C}}_{f^q}^*\right]\,,
 \end{eqnarray}
\noindent which will be our starting point for the analysis of the effect of the final state interaction. Recalling that the S-matrix is unitary by definition and its elements are an overlap between in and out states, Eq.~\eqref{cpuspin} can be rewritten as~ \cite{wolfenstein,BediagaPRD2014}  (see also \ref{ap:cpts}):
\begin{small}
\begin{equation}\label{cpuspin-5}
\Delta\Gamma_{CP}(f^{q})
 =
 2\,\text{Im}[V_{ub}^*V_{uq}V_{cb}V^*_{cq}]
 \hspace{-.2cm}\sum_{f^{\prime q},f^{\prime\prime q}}\hspace{-.2cm}\text{Im}\Big[S_{f^{\prime q}f^{q}}
 S^*_{f^{\prime\prime q}f^{q}}\,\left\{
\mathcal{U}_{f^{\prime q}}^*
 \mathcal{C}_{f^{\prime\prime q}}+
\bar{\mathcal{U}}_{f^{\prime q}}^* 
\bar{\mathcal{C}}_{f^{\prime\prime q}}
 \right\} \hspace{-.1cm}\Big]\,.
 \end{equation} 
 \end{small}
This is our main formula, exposing explicitly the effect of the FSI and the CP-violating phase for the decay channels with  $\Delta S=1$ and $\Delta S=0$ carrying different
net  strangeness, and therefore not coupled by the strong interaction. 

The CP-violating phase enters linearly at the lowest order in the hadron decay amplitude. If we impose CPT invariance of the strong sector, independently of the weak Hamiltonian,  using the steps given in Refs.~\cite{BediagaPRD2014,wolfenstein} and summarized in the \ref{ap:cpts}, it is easy to  show that 
 the sum over intermediate channels $f^q$  in $\Delta\Gamma_{CP}$ gives:
\begin{equation}\label{cptpartial-1}
   \sum_{f^q} \Delta\Gamma_{CP} ( f^q)=2\,\text{Im}[V_{ub}^*V_{uq}V_{cb}V^*_{cq}]\,\sum_{f^q}\text{Im}\left[\mathcal{U}_{f^q}\, \mathcal{C}_{f^q}^*+\bar{\mathcal{U}}_{f^q}\, \bar{\mathcal{C}}_{f^q}^*
\right]=0\, .
\end{equation} 
  This zero is then a direct consequence of the CPT constraint to channels coupled by the strong final state interaction that we named as sCPT. Therefore, the FSI should bring phases that are compensated by the different signs of $\Delta_{CP}(f^q)$ for channels coupled by the strong interaction.
 The form of sCPT relation though is more restrictive than the one written in Eq.~\eqref{cpt5a}, that could be also derived from the unitarity constraint of the CKM matrix~\eqref{eq:uspin0-1} and U-spin symmetry relation~\eqref{h-hbaruspin}.

 Finally, one can identify two sources of  the global CP asymmetry signs: the weak U-spin symmetry between final states with different strangeness and the CPT constraint between states coupled by the strong interaction with the same quantum numbers.

\section{Coupled $\pi\pi$ and $KK$ channels in  $B^\pm$ three-body  decays}

In Ref.~\cite{BediagaPRD2014,Nogueira2015}  we had discussed the role of the coupling between $\pi\pi \to KK$ as a mechanism to explain the  total $\acp$ observed in the charged three-body B decays. Here, we recall this argument with the formalism developed above.
 In order to apply to the $B^ \pm$ decay channels
coupled by the strong interaction,  we start by naming them to stress the strangeness and the pair of mesons that will couple through FSI:   $d_{\pi\pi}\equiv\pi^\pm \pi^ +\pi^-$ and $d_{KK}\equiv\pi^\pm K^ +K^-$,
$s_{\pi\pi}\equiv K^\pm \pi^+ \p^-$ and $s_{KK}\equiv K^\pm K^+K^-$ . In such case,  the pairs of
coupled channels are $\{d_{\pi\pi},d_{KK}\}$ and $\{s_{\pi\pi},s_{KK}\}$, which interact via rescattering between $\pi\pi$ and  $KK$.

Considering only the interaction in S-wave 
the two-body,  $\pi\pi$ and $KK$, coupled-channel S-matrix is:
\begin{equation}\label{eq:s00}
\begin{pmatrix}
S_{\pi\pi,\pi\pi} &  S_{\pi\pi,K\bar K} \\
 S_{K\bar K,\pi\pi} & S_{K\bar K,K\bar K}
\end{pmatrix}=
\begin{pmatrix}
\eta \,\text{e}^{2\imath\delta_{\pi\pi}} &  \imath \sqrt{1-\eta^2}\,\text{e}^{\imath (\delta_{\pi\pi}+\delta_{KK})} \\
 \imath \sqrt{1-\eta^2}\,\text{e}^{\imath (\delta_{\pi\pi}+\delta_{KK})} & \eta\, \text{e}^{2\imath\delta_{KK}} 
\end{pmatrix}\, ,
\end{equation}
where $\delta_{\pi\pi}$ and $\delta_{KK}$ are the phase-shifts, $1\ge\eta\ge 0$ is the absorption parameter. 

In the leading order (LO) of the strong interaction~\cite{BediagaPRD2014}, namely taking into account the transition matrix at the lowest order in Eq.~\eqref{cpuspin-5}, and identifying for the two-channel case $q_1\equiv q_{\pi\pi}$ and $q_2\equiv q_{KK}$, for $q= s$ or $d$, we can find that:
\begin{equation}
\label{cpuspin-6LO}
\begin{aligned}
\Delta\Gamma^{(\text{LO})}_{CP}( q_{\pi\pi})= 
w_q 
 \,\,\text{Re}\,\Big[& 
  \,\text{e}^{\imath (\delta_{\pi\pi}-\delta_{KK})}\,
\left\{\mathcal{U}_{0q_{\pi\pi}}^*
 \mathcal{C}_{0q_{KK}}+\bar{\mathcal{U}}_{0q_{\pi\pi}}^*
 \bar{\mathcal{C}}_{0q_{KK}} \right\}
 \\ &-\text{e}^{-\imath (\delta_{\pi\pi}-\delta_{KK})} 
\,
\left\{\mathcal{U}_{0q_{KK}}^*
\ \mathcal{C}_{0q_{\pi\pi}}+\bar{\mathcal{U}}_{0q_{KK}}^*
\ \bar{\mathcal{C}}_{0q_{\pi\pi}}\right\}
 \Big] 
 \,,
\end{aligned}
 \end{equation}
where 
$w_q= 2\eta\sqrt{1-\eta^2}\,\,\text{Im}[V_{ub}^*V_{uq}V_{cb}V^*_{cq}]\,$. Note that  we imply due to the CPT relation that $\mathcal{U}_{0q_{\pi\pi}}\mathcal{C}_{0q_{\pi\pi}}^*$, $\mathcal{U}_{0q_{KK}}\mathcal{C}_{0q_{KK}}^*$  and the analogous products for the conjugate states are real, as they do not contain the distortion from the FSI. This assumption simplifies the partial width difference between the two charge conjugated decays. The result shows that in  LO only the interference between  S-matrix off-diagonal elements in~\eqref{eq:s00}  contribute to  $\Delta\Gamma_{CP}(q_i)$.

The U-spin symmetry within this example corresponds to:
\begin{eqnarray}
 &&   \mathcal{U}_{0d_{\pi\pi}}=\mathcal{U}_{0s_{KK}}\quad\text{and}\quad \mathcal{U}_{0d_{KK}}=\mathcal{U}_{0s_{\pi\pi}}\, ,
 \nonumber \\
  &&   \mathcal{C}_{0d_{\pi\pi}}=\mathcal{C}_{0s_{KK}}\quad\text{and}\quad \mathcal{C}_{0d_{KK}}=\mathcal{C}_{0s_{\pi\pi}}\, .\,
  \label{eq:UspinLO}
\end{eqnarray}
 and the analogous relations for the amplitudes of the charge conjugate states.

As we argue in section \ref{hfsi}, after integrating $\Delta\Gamma_{CP}(q_i)$ over the phase-space, only off-diagonal channels will survive and contribute to the global CPV. 
In addition, if we assume $\delta_{\pi\pi}\approx\delta_{KK}$ and equal masses for the pion and kaon, which means the FSI does not distinguish the change of $\pi\leftrightarrow K$, and 
 taking into account the opposite signs in $w_d=-w_s$, from the unitarity of the CKM matrix, we get that: 
\begin{equation}\label{uspinlo2ch-1}
 \frac{\Delta\Gamma_{CP}(\pi^\pm K^+ K^-)}{\Delta\Gamma_{CP}(K^\pm \pi^+ \pi^-)}\sim -1\,
  \quad\text{and} \quad
 \frac{\Delta\Gamma_{CP}(\pi^\pm \pi^+ \pi^-)}{\Delta\Gamma_{CP}(K^\pm K^+ K^-)}\sim -1\,.
\end{equation}
From the sCPT  relation
$\Delta\Gamma({q_{\pi\pi}})= -\Delta\Gamma({q_{KK}})$ in
Eq.~\eqref{cpuspin-6LO}, we get that
\begin{equation}\label{uspinlo2ch-2}
\frac{\Delta\Gamma_{CP}(\pi^\pm K^+ K^-)}{\Delta\Gamma_{CP}(\pi^\pm \pi^+ \pi^-)}= -1 \quad\text{and} \quad \frac{\Delta\Gamma_{CP}(K^\pm K^+ K^-)}{\Delta\Gamma_{CP}(K^\pm \pi^+ \pi^-)}= -1\, 
\end{equation}
The first ratio  in Eq.~\eqref{uspinlo2ch-1}
is consistent with 
what was predicted by U-spin symmetry and  with Eq.~\eqref{eq:deltacpexp-a}, as we have already discussed.
The first and second theoretical ratios given in  Eq.~\eqref{uspinlo2ch-2} with values of -1 are compatible within $1\sigma$ with the experimental  ratios of $-0.73\pm 0.22$ and  $-0.81\pm 0.31$, respectively. 
We remind that these ratios were obtained  from Table~\ref{W_cp}, which was built with  the available experimental data for the $B$ decay rates and CP asymmetry. We stress that within a two coupled-channel picture the ratios~\eqref{uspinlo2ch-1} and~\eqref{uspinlo2ch-2} are valid beyond the LO,  and due to that the superscript (LO) was dropped out  in those equations. 

It is important to be aware of the approximations imposed in the above calculations.  The assumption of U-spin symmetry under the transformation $\pi\leftrightarrow K$ as expressed  by the relations~\eqref{eq:UspinLO},   in addition to the equality between the phase-shifts for the elastic $\pi\pi$ and $KK$ channels and masses,
can affect the magnitudes of the global CPV. Thus, we should have caution  when comparing the magnitudes in the ratios given by Eq.~\eqref{uspinlo2ch-1} with the experimental data, while its sign is well defined.
With respect to the relations~\eqref{uspinlo2ch-2}, they are strongly grounded in: (i) the experimental observation  of the global CPV signal originated from the kinematic region where $\pi\pi\to KK$ scattering  is dominant, 
and (ii) the CPT relation considering only the coupling between these two channel. These considerations are indeed supported by the comparison between the experimental values of $A_{CP}$'s in the two last columns of Table~\ref{ACP_rescattering}. Finally, the complementary experimental ratios given in Eq.~\eqref{eq:deltacpexp-b} can be understood within $1\sigma$ as a direct consequence of combining the theoretical ratios from Eqs. \eqref{uspinlo2ch-1} and \eqref{uspinlo2ch-2}.

\section{Final Remarks}

 Our study shows the relevance of the FSI to the global CPV in $B^\pm\to h^\pm h^+h^-$ addressed by the ratio of charge conjugate width differences and given by~\eqref{uspinlo2ch-2}.    
The comparison of our results with the experimental values in Eqs.~\eqref{eq:deltacpexp-a} and \eqref{eq:deltacpexp-b},  stresses that the used  U-spin symmetry  at the hadronic level, namely, the exchange $K\leftrightarrow \pi$
  in $B^\pm$ decay channels are supported by the data. 
 
The proposed form to apply the U-spin symmetry, together with the sCPT constrain including the FSI, can reveal the correct relative signs between the $\Delta\Gamma_{CP}$'s of the charged three-body $B$ decays,  as one sees by comparing the ratios \eqref{uspinlo2ch-1} and \eqref{uspinlo2ch-2}, with those extracted from the experimental values presented in Eqs.~\eqref{eq:deltacpexp-a} and \eqref{eq:deltacpexp-b}. Note that the magnitudes are reproduced within the experimental errors.

Although data is still not as precise as we would desire,  there will be new high statistics in the near future by LHCb (Run 2 and Run 3) and from Belle2  which  will allow us to better address this issue. From the theoretical side, in the proposed CPT constrained framework including FSI, we only take into account the S-matrix in the charged coupled channels $\pi\pi$ and $KK$ in the S-wave. But besides the interactions among the charged mesons, one can have the coupling to the neutral ones along with other isospin zero  meson pairs such as $\eta\eta$ as discussed in detail in~\cite{pat-mane2020}.  
It was shown by many theoretical studies including the recent one~\cite{pat-mane2020} that $KK$ coupling to $\pi\pi$ channel is enhanced in the S-wave by the superposition of resonance $f_0(980)$ just before the $KK$ threshold. The coupling between this two channels is needed for the theoretical description of the $\pi\pi\to \p\p$ experimental scattering data.
The situation with $\eta\eta$ is different as it is not strongly coupled to the $\pi\pi$ and $KK$ channels.

If we consider the  coupled-channel contributions from the neutral mesons as well, we will be able to expand the B three-body decays that we can connect through FSI. Indeed, for $\Delta S=0$, involving kaons and pions there are: 
\begin{equation}\label{eq:5ch-ds0}
 B^\pm \to \pi^\pm K^+ K^-,\,\,
 \pi^\pm K^0\bar K^0,\,\, 
K^\pm\bar K^0\pi^0,\,\,
 \pi^\pm\pi^+\pi^-,\,\, \pi^\pm\pi^0\pi^0.
 \end{equation}
And another  five channels with the same characteristic for $\Delta S  =1$:
\begin{equation}\label{eq:5ch-ds1}
B^\pm \to K^\pm\pi^+\pi^-,\,\,
\pi^\pm K^0 \pi^0,
  K^\pm\pi^0\pi^0,\,\,
 K^\pm K^0\bar K^0,\, \, K^\pm K^+K^-.
 \end{equation}

The formula we wrote for $\Delta\Gamma_{CP}$ given by Eq.~\eqref{cpuspin-5} is general and can incorporate those couplings, which will change the magnitude of the ratios~\eqref{uspinlo2ch-1} and~\eqref{uspinlo2ch-2}, but we expect not the relative signs.
The sCPT relation given by Eq.~\eqref{cptpartial-1} allows us
to write down the relation between the $\Delta\Gamma_{CP}$'s, independently for the five decay channels with $\Delta S=0$ \eqref{eq:5ch-ds0} and $\Delta S=1$ \eqref{eq:5ch-ds1}.
Furthermore, we expect that the channels $K^\pm \bar K^0 \pi^0$ and $\pi^\pm K^0 \pi^0$ are weakly coupled to the other four channels with $\Delta S=0$ and $\Delta S=1$, respectively,  as the three-body rescattering that couple these two states with the other four should be suppressed, as it requires two-loop processes~(see e.g. \cite{NogueiraFBS2017}).
Removing them, we have for $\Delta S  = 0$: 
\begin{small}
\begin{equation}
 \Delta\Gamma_{CP}(\pi^\pm K^+ K^-) + 
 \Delta\Gamma_{CP}(\pi^\pm K^0\bar K^0) 
 + \Delta\Gamma_{CP}(\pi^\pm\pi^+\pi^-)+ 
\Delta\Gamma_{CP}(\pi^\pm\pi^0\pi^0) =0\, .
\end{equation}
\end{small}
The other sCPT equation for $\Delta S  =1$ is given by: 
\begin{small}
\begin{equation}
 \Delta\Gamma_{CP}(K^\pm \pi^+\pi^-) 
 +\Delta\Gamma_{CP}(K^\pm
 \pi^0\pi^0) + %\nonumber\\
 \Delta\Gamma_{CP}(K^\pm K^+ K^-)+
\Delta\Gamma_{CP}(K^\pm K^0 \bar K^0) =0.
\end{equation}
\end{small}

It is also reasonable to expect that the charged and noncharged channels have similar decay amplitudes, and for $\Delta S=0$ we have that:
\begin{equation}\label{eq:sCPT-1}
 \frac{\Delta\Gamma_{CP}(\pi^\pm K^+ K^-)}{
 \Delta\Gamma_{CP}(\pi^\pm K^0\bar K^0)}\sim 1 \quad\text{ and }\quad
\frac{ \Delta\Gamma_{CP}(\pi^\pm\pi^+\pi^-)}{
\Delta\Gamma_{CP}(\pi^\pm\pi^0\pi^0)}\sim 1\, ,
\end{equation}
and for $\Delta S  =1$: 
\begin{equation} \label{eq:sCPT-2}
\frac{ \Delta\Gamma_{CP}(K^\pm \pi^+\pi^-)}{
 \Delta\Gamma_{CP}(K^\pm
 \pi^0\pi^0)}\sim 1\quad\text{ and }\quad
\frac{ \Delta\Gamma_{CP}(K^\pm K^+ K^-)}{
\Delta\Gamma_{CP}(K^\pm K^0 \bar K^0)}\sim 1.
\end{equation}

Making use of our relations for the ratios of  \CP asymmetry partial widths, Eqs.~\eqref{uspinlo2ch-1} and \eqref{uspinlo2ch-2}, in addition to the approximate relations \eqref{eq:sCPT-1} and \eqref{eq:sCPT-2}, we can predict that:
\begin{equation}\label{uspinlo2ch-20}
\frac{\Delta\Gamma_{CP}(\pi^\pm K^0 \bar K^0)}{\Delta\Gamma_{CP}(\pi^\pm \pi^0 \pi^0)}\sim -1 \quad\text{and} \quad \frac{\Delta\Gamma_{CP}(K^\pm K^0 \bar K^0)}{\Delta\Gamma_{CP}(K^\pm \pi^0 \pi^0)}\sim -1\, .
\end{equation}

With the above equations and the already observed experimental results for the charged modes, we can make predictions for the neutral channels yet without experimental results.  
The LHCb upgrade together with the Belle II, that is already taking data, can give us  a good experimental estimate of the different CP asymmetries. 
These future experimental data will provide further support to the proposed theoretical framework to describe the global CP violation in charmless three-body $B$ decays,  which unifies  U-spin symmetry and  final state interactions
at the hadronic level within a CPT invariant approach.

{\it Acknowledgments.} 
We would like to thank Jussara Miranda for the fruitful discussions.
This study was financed in part by Conselho Nacional de Desenvolvimento Cient\'{i}fico e Tecnol\'{o}gico (CNPq) under the grant 308486/2015-3 (TF) and INCT-FNA project 464898/2014-5,   FAPESP Thematic Projects grant   2017/05660-0 and 2019/07767-1 (TF), FAPERJ, CAPES and  CAPES - PRINT grant 88887.580984/2020-00. PCM would like to thank University of Bristol for the support as well.
 
\appendix

\section{Strong CPT relation}
\label{ap:cpts}

In this Appendix,  we follow Ref.~\cite{BediagaPRD2014} and sketch the derivations of Eq.~\eqref{cpuspin-5} from \eqref{cpuspin} and  the resulting  relation \eqref{cptpartial-1} expressing the sCPT constraint.
The requirement of \CPT invariance for the weak Hamiltonian is fulfilled by the matrix element of the decay amplitude when~\cite{bigi_book}:
\begin{eqnarray}
&&\langle  f_{out}|\mathcal{H}_{\text{w}} |B^u\rangle
= \chi_B\chi_ f\langle  \bar  f_{in} |\mathcal{H}_{\text{w}} | \bar B^u  \rangle^*\, ,\label{cpt3}
 \end{eqnarray}
 where $\chi_B$ and $\chi_ f$ are constant phases. Taking into account the decomposition  of \eqref{cpt3} in terms of the matrix elements of the operators $U^q$ and $C^q$ given in Eq.~\eqref{cpt3eff0}, and the charge conjugate operators $\bar U^q$ and $\bar C^q$ present in Eq.~\eqref{eq:barBbarf0}, one finds that:
\begin{equation}
\mathcal{U}_{f^q}=\langle f^q_{out}|U^q| B^u  \rangle =  \chi_B\chi_{ f^q}\langle  \bar  f^q_{in} |\bar U^q| \bar  B^u  \rangle^*
\, .\label{cp11}
\end{equation}
and  the analogous relations for $\mathcal{C}_{f^q}$ corresponding to the matrix elements of the operator $C^{q}$.
If we consider: (i) the unity resolution in terms of the $| f^q_{out}\rangle$ states; and (ii)  the strong  S-matrix element fulfill 
$ S_{ f^\prime f}=
\langle \bar  f^\prime_{out}|\bar  f_{in}\rangle
=\langle   f^\prime_{out}|  f_{in}\rangle \,
$; 
one easily arrives to:
\begin{eqnarray}\label{cp110}
\mathcal{U}_{f^q}=\chi_B\chi_{ f^ q}\sum_{ f^{\prime q}} S_{ f^{\prime q} f^ q}\langle  f^{\prime q}_{out} | U^q   | B^u \rangle ^*=\chi_B\chi_{ f^ q}\sum_{ f^{\prime q}} S_{ f^{\prime q} f^ q}\,\mathcal{U}_{f^{\prime q}}^*\, ,
 \end{eqnarray}
 and an analogous relation for the matrix elements of $C^{q} $,  and for the matrix elements of the charge conjugate states.  Therefore, our CP asymmetry expression, Eq.~\eqref{cpuspin-5}, namely, the difference of partial widths of charge conjugate states follows
 from Eqs.~\eqref{cp110}, \eqref{cpt3eff0} and \eqref{eq:barBbarf0}.
 
One can verify the implication of the \CPT symmetry imposed in  Eq.~\eqref{cpuspin-5} by summing 
over channels coupled by the strong interaction and making 
use of the S-matrix unitarity, which leads to:
\begin{equation}\label{cptpartial-2}
\sum_ {f^q}\Delta\Gamma_{CP}(f^q)
 =
 2\,\text{Im}[V_{ub}^*V_{uq}V_{cb}V^*_{cq}]\,\sum_{ f^q}\text{Im}\left[\mathcal{U}_{f^{ q}}^*\mathcal{C}_{f^{ q}}+\bar{ \mathcal{U}}_{f^{ q}}^*\bar{\mathcal{C}}_{f^{ q}}
 \right]\,,
 \end{equation} 
with this sum being equal to the one written in Eq.~\eqref{cptpartial-1}, and then:
\begin{equation}\label{cptpartial-3}
\sum_{f^q}\text{Im}\left[ \mathcal{U}_{f^q}^* \mathcal{C}_{f^q}+\bar{\mathcal{U}}_{f^q}^* \bar{\mathcal{C}}_{f^q}\, \right] =
\sum_{f^q}\text{Im}\left[\mathcal{U}_{f^q}\, \mathcal{C}_{f^q}^*+\bar{\mathcal{U}}_{f^q}\, \bar{\mathcal{C}}_{f^q}^* \right]
=0\,,
 \end{equation} 
what proves Eq.~\eqref{cptpartial-1} constraint, namely the sCPT relation.

%%%----------------------%%%
%%%%%% BIBLIOGRAPHY %%%%%%%%
%%%----------------------%%%

\end{document}